\documentclass[%
 reprint,
 amsmath,amssymb,
 aps,
]{revtex4-2}

\usepackage{graphicx}
\usepackage{parskip} 
\usepackage{physics}
\usepackage{dcolumn}
\usepackage{bm}
\usepackage{amsfonts,amssymb,amsbsy, amsmath}
\usepackage{xcolor}

\begin{document}

\preprint{APS/123-QED}

\title{Observation of the two-photon Landau-Zener-St\"uckelberg-Majorana effect}

\author{Isak Bj\"orkman}
\affiliation{InstituteQ and QTF  Centre  of  Excellence,  Department  of  Applied  Physics, School  of  Science,  Aalto  University,  FI-00076  Aalto,  Finland}
\author{Marko Kuzmanovi\'c}
\affiliation{InstituteQ and QTF  Centre  of  Excellence,  Department  of  Applied  Physics, School  of  Science,  Aalto  University,  FI-00076  Aalto,  Finland}
\author{Gheorghe Sorin Paraoanu}\thanks{sorin.paraoanu@aalto.fi}
\affiliation{InstituteQ and QTF  Centre  of  Excellence,  Department  of  Applied  Physics, School  of  Science,  Aalto  University,  FI-00076  Aalto,  Finland}

\date{\today}

\begin{abstract}
Second-order processes introduce nonlinearities in quantum dynamics, unlocking a totally unexpected area of control operations. Here we show that the well-known Landau-Zener-St\"uckelberg-Majorana (LZSM) transition can be driven by a virtual process in a three-level system whereby two photons from a drive with linearly-modulated phase create excitations onto the third level while avoiding completely the first level. We implement this experimentally in a transmon qubit achieving a population transfer of $98\%$, limited by relaxation. We predict and observe experimentally the doubling of the LZSM velocity.
The observation of this effect is made possible by the nearly-exact cancellation of the two-photon ac Stark shift when the third transition is included.
Furthermore, we demonstrate considerable robustness to offsets in frequency and amplitude, both in theory and experimentally. 
\end{abstract}

\maketitle


In quantum mechanics, perturbative effects can be described by expanding in power series in the interaction coupling, revealing higher-order processes that can be systematically acounted for using for example Feynman diagrams. However, this methodology cannot be applied for non-perturbative processes, of which tunneling is an outstanding  example. Here we address this problem for a specific form of tunneling, the paradigmatic 
Landau-Zener-St\"uckelberg-Majorana (LZSM) effect known for more than 90 years since the seminal four papers from 1932 \cite{Landau1932,Zener1932,Stueckelberg1932,Majorana1932}. LZSM tunneling appears in a variety of problems: the quantum Ising model driven through the critical point at a finite rate \cite{Dziarmaga2005}, in the Kibble-Zurek theory of the topological defect production in nonequilibrium  second order phase transitions \cite{Damski2005},
and in adiabatic quantum computing \cite{Rose2009}. It also underlies relativistic effects such as the Klein tunneling through barriers created by electric fields \cite{Solano2010}, or the Schwinger effect and its realization as Landau-Zener breakdown in band insulators under a strong electric field \cite{Aoki2014}.
  
The single-photon Landau-Zener transfer can be realized by modulating linearly the transition frequency of a qubit near an avoided crossing. This leads to a Hamiltonian of the general form
\begin{equation}
	H_{\rm LZSM} = -\frac{\hbar}{2} (vt - \delta )\sigma^{z} + \frac{\hbar \Omega}{2} \sigma^{x},
\end{equation}
where $\delta$ is a fixed detuning and $v$ is the LZSM speed.
If the proceess starts at $-T/2$ and ends at $T /2$ and if $|\delta /v| \ll T$, the celebrated LZSM formula 
\begin{equation}
	P_{\rm LZSM} = \exp\left[-\pi \frac{\Omega^2}{2|v|}\right]
\end{equation}
provides an excellent approximation to the nonadiabatic transition probability \cite{Shevchenko2010, Silveri_2017, Shevchenko_2023}.

It is not obvious how to obtain higher-order processes derived from LZSM effect, as this process is manifestly non-perturbative. A starting observation is that, since virtual processes have relatively small probabilities, in order to observe a second-order process one must look for a situation where the first order effect is small or even zero. This situation occurs naturally in weakly anharmonic systems such as the transmon, where the direct transition between the ground state and the second excited state is forbidden by a selection rule. Thus, driving a virtual-excitation process between these states may be the key to realize a second-oder LZSM. In particular, we will show that for such a process the hallmark is that the LZSM speed is twice as large as that of the standard single-photon effect. The equivalence between the two-photon driven multilevel transmon and the two-level LZSM Hamiltonian is possible due to the fortuitous cancellation of the two-photon ac Stark shift by a contribution due to the coupling of the drive with the third excited state.

Besides illuminating a fundamental conundrum in quatum physics, applications of the two-photon LZSM include situations where direct coupling is either undesired or prohibited due to selection rules - for example state transfer between distant nodes without populating the intermediate states \cite{Cleland2020}, controlled emission of photons in cavities \cite{Bergmann_2019}, detection of virtual photons in ultrastrongly coupled systems \cite{Carusotto_2012,Huang_2014,Falci_2017,Giannelli_2023,Giannelli_2024}, and quantum information processing with qudits \cite{Sanders_2020,Siddiqi_2021, Siddiqi_randomized_2021, Guzik_2022,Yu_2023, Schuster_2023}. Due to the fact that the modulation depth is of the order of only 10 MHz, the method is suitable for population transfer in frequency-crowded systems.  Also, two-photon processes have become an important method in dissipation engineering for state stabilization \cite{puri2017engineering, grimm2020stabilization, berdou2023one, Leghtas2015}, already  leading to applications such as single-photon microwave detectors \cite{Flurin2020}.

For the experimental proof of concept we use a superconducting transmon with $E_\mathrm{C} \approx h \times 340$ MHz and a Josephson energy of $E_\mathrm{J} \approx h \times 21.4$ GHz, yielding $E_\mathrm{J} /E_\mathrm{C} \approx 62.9$. The transition frequencies are $\omega_{\rm ge}/2 \pi = 7.24$ GHz and $\omega_{\rm ef}/ 2\pi =  6.90$ GHz, and the drive pulse duration is fixed to  $T = 400$ ns. We drive the transmon with a constant $LO = 7.0$ GHz signal mixed in an IQ-mixer with a frequency modulated $IF(t)$ signal. The up-converted sideband is suppressed by tuning the relative phase and amplitude of the I and Q quadratures, ultimately leaving a single-sideband signal at $LO-IF(t)$, limited by the $\pm 350$ MHz bandwidth of the $IF$-signal generator.

When the transmon is driven by a microwave field with frequency $\omega_{d}$, in a frame co-rotating with the drive and by applying the rotating wave approximation, we have
\begin{equation}
	\begin{aligned}
		H = -\hbar \Delta_{\rm ge} \ket{\mathrm{g}}\bra{\mathrm{g}} + \frac{\hbar\Omega_{\rm ge}}{2}\big(\ket{\mathrm{g}}\bra{\mathrm{e}} + \ket{\mathrm{e}}\bra{\mathrm{g}}\big) \\ + \frac{\hbar \Omega_{\rm ef}}{2}\big(\ket{\mathrm{e}}\bra{\mathrm{f}} + \ket{\mathrm{f}}\bra{\mathrm{e}}\big) + \hbar\Delta_{\rm ef} \ket{\mathrm{f}}\bra{\mathrm{f}} .
	\end{aligned}	
	\label{rotHamiltonian}
\end{equation}
Here  $\Delta_{\rm ge} = \omega_{\rm ge}-\omega_d - (\partial \omega_d /\partial t) t$ and $ \Delta_{\rm ef} = \omega_{\rm ef}-\omega_d - (\partial \omega_d/\partial t ) t$ are the detunings of the drive frequencies from the two transitions.
In the following we denote for simplicity $\Omega_{\rm ge}=\Omega$, while $\Omega_{\rm ef}$ is rescaled proportionally with the ratio of the effective electric dipole moments of the two transitions $\Omega_{\rm ef}=g_{ef}/g_{ge}\Omega=\sqrt{2}\Omega$ \cite{koch2007charge}. The selection rules prohibit a direct transition from $\ket{g}$ to $\ket{f}$, hence no $\ket{g}\bra{f}$-term appears.

We drive the system during the time interval $t \in [-T/2, T/2]$ with a constant chirp rate $-v/2$, but this time around the 
$\omega_{\rm gf}/2 = (\omega_{\rm ge}  + \omega_{\rm ef})/2$,
\begin{equation}
	\omega_{d}(t) = \omega_{\rm gf}/2 - vt/2 + \delta .
	\label{driveformd}
\end{equation}
Here the value of $v$ together with $\tau$, the pulse duration, sets the modulation depth $D=\omega_{d}(T/2) - \omega_{d}(-T/2) = -vT/2$, while $\delta$ is a fixed offset from half the $\rm gf$-transition frequency.

This leads to
\begin{align}
	\Delta_{\rm ge}(t) = \omega_{\rm ge}-\omega_{\rm gf}/2 + vt - \delta = \Delta-\delta_{\rm gf}(t),
	\label{delta1201}\\
	\Delta_{\rm ef}(t) = \omega_{\rm ef}-\omega_{\rm gf}/2 + vt - \delta = -\Delta-\delta_{\rm gf}(t) 
	\label{delta0112}.
\end{align}
Experimentally $\Delta = \omega_{\rm ge}- \omega_{\rm gf}/2 = \omega_{\rm gf}/2 - \omega_{\rm ef} \approx  E_{\rm C}/2 \hbar$ is set by the anharmonicity of the transmon, and $\delta_{\rm gf}(t) = -v t + \delta$ is the detuning with respect to $\omega_{\rm gf}/2$.

As with the usual LZSM process, the transfer from $\ket{g}$ to $\ket{f}$ is realized by adiabatically following one of the eigenstates of the drive Hamiltonian.

However, this cannot be realized by a pulse with a constant amplitude as none of the instantaneous eigenstates start in the ground state nor end in the second excited state exactly, but rather in a combination of all three levels. 
To avoid this, we modulate the amplitude of the pulse such that $\Omega(t = \pm T/2) \approx 0$. In this way, the Hamiltonian in Eq. \eqref{rotHamiltonian} is diagonal at the beginning and at the end, meaning that $\ket{\mathrm{g}}$ is an eigenstate of the Hamiltonian at $t = -T/2$. Thus we eliminate spurious Rabi flopping, which is critical for achieving robustness of the transfer. These criteria are approximately met with a super-Gaussian envelope $\Omega(t) \propto \exp\left[K_c(2t/T)^{n}\right]$, where $K_c$ is the cut-off value and $n$ is the order. Here we chose $K_c = \ln(0.01)$ and $n = 4$.

\begin{figure}[h!]
	\centering
	\includegraphics[scale=0.39]{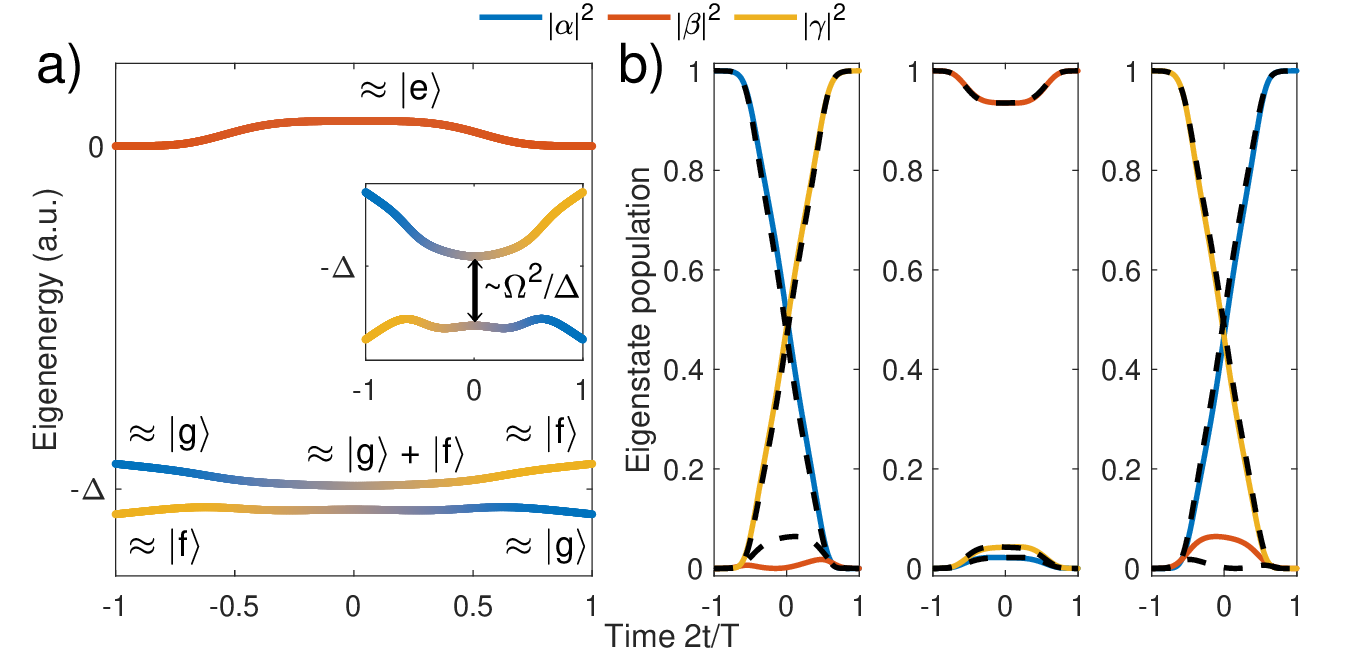}
	\caption{(a) Instantaneous eigenenergies of the Hamiltonian in Eq. \eqref{rotHamiltonian} for $v>0$, $\delta =0$, and for a super-Gaussian envelope: the eigenenergies are shown in colors according to the composition of its eigenstate. (b) Instantaneous eigenstates: each panel is one eigenstate constructed as a linear combination of the states $\ket{g}$, $\ket{e}$, and $\ket{f}$. The black dashed lines are the instantaneous eigenstates for the drive having a positive chirp rate ($v<0$).}
	\label{ieigp5}
\end{figure}

Each of the three instantaneous eigenenergies of the Hamiltonian in Eq. \eqref{rotHamiltonian} 
can be expanded in the qutrit basis in the general form  $\alpha \ket{\mathrm{g}} + \beta \ket{\mathrm{e}} + \gamma \ket{\mathrm{f}}$. In Fig. \ref{ieigp5} we present them with lines colored based on the weight of each component in the eigenstate: ground state is blue, the first excited state is red and the second excited state is yellow. When the eigenstate composes several states, the colors mix according to the composition.

In Fig. \ref{ieigp5} a), we find three eigenenergies: one near zero and two close to $-\Delta$. The eigenenergy near zero, has an eigenstate that mainly constitutes of the first excited state, where the shift from zero is the AC Stark effect \cite{autler1955stark}. The two remaining eigenstates, at approximately $-\Delta$, initially constitute the ground state and the second excited state. When on-resonance with $\omega_{\rm gf}/2$, these eigenstates are a superposition of the ground state and the second excited state and their wavefunctions exchange character. By the end of the drive, the $\ket{g}$ and $\ket{f}$ components have interchanged completely, finalizing the LZSM process. 
Depending on the sign of the modulation, i.e.whether we approach $\omega_{rm gf}/2$ from above or below, the instantaneous eigenstates have slightly different character: for a negative modulation depth the eigenstate which initially corresponds to the ground state does not posses a significant $\ket{e}$ contribution, while the one which is initially $\approx \ket{f}$ does. The effect is reversed by changing the sign of the modulation, which is also shown in Fig. \ref{ieigp5}. Therefore, a drive pulse with a negative modulation depth is preferred for transferring the population from $\ket{g}$ to $\ket{f}$.

According to \cite{james2007effective}, under the assumptions (i) $\omega_{\rm ge}-\omega_d, \omega_d -\omega_{\rm ef} > 0$ $\forall$ $t$, (ii) a possible change in the drive frequency over time is slow in comparison to everything else, and (iii) the interaction is weak enough such that fourth and higher order effects can be neglected, an effective low-frequency Hamiltonian can be obtained. This results in $\ket{\mathrm{g}}$ and $\ket{\mathrm{f}}$ having an effective coupling proportional to $\Omega^2/\Delta$, through a second-order process responsible for Raman-like transitions \cite{james2007effective}. An additional shift is present due to the fourth energy level, $\ket{\rm h}$: overall, this makes the total shift in the transition frequency between $\ket{\rm g}$ and $\ket{\rm f}$ as $\frac{\Omega^2}{4(\Delta-\epsilon_{\rm gf})}-\frac{\Omega^2_{\rm ef}}{4(\Delta+\epsilon_{\rm gf})}+\frac{\Omega^2_{\rm fh}}{4(3\Delta+\epsilon_{\rm gf})}$, where $\epsilon_{\mathrm gf}(t) = D t/T + \delta$ is the offset in drive frequency from $\omega_{\rm gf}/2$. Here $\Omega_{\rm fh}$ is the coupling of the drive with the third transition, scaling proportional to the electric dipole $\Omega_{\rm fh} = \sqrt{3} \Omega$ \cite{koch2007charge}. For our experimental parameters $\epsilon_{\rm gf}(t)$ is small compared to $\Delta$ and therefore  $\Delta \pm \epsilon_{\rm gf}(t) \approx \Delta$, resulting in the cancellation of the ac Stark shift. 

 Overall, this results in the effective Hamiltonian
\begin{equation}
	\begin{aligned}
		H_{\rm eff} \approx 
		& \frac{\hbar}{2} \left( 
		2\epsilon_{\rm gf} + 2t\frac{d \epsilon_{\rm gf}}{dt}
		\right) (|\mathrm{f} \rangle \langle \mathrm{f}| - |\mathrm{g} \rangle \langle \mathrm{g}|)  \\
		&-\frac{\hbar \Omega_{\rm ge} \Omega_{\rm ef}}{4 \Delta} (|\mathrm{g} \rangle \langle \mathrm{f}| + |\mathrm{f} \rangle \langle \mathrm{g}|)	
	\end{aligned} 	\label{eigenstHam}
\end{equation}

which describes the two-photon Landau-Zener transfer

\begin{equation}
	\begin{aligned}
		H_{\rm LZSM}^{(\rm 2ph)} \approx 
		 \frac{\hbar}{2} \left( 
		-2 vt +2\delta 
		\right)  \sigma^{z}_{\rm gf} 
		-\frac{\hbar \Omega_{\rm 2ph}}{2} \sigma^{x}_{\rm gf}
	\end{aligned} 	\label{eigenstHam2}
\end{equation}

where $\sigma^{z}_{\rm gf}= |f \rangle \langle f| - |g \rangle \langle g|$
and $\sigma^{x}_{\rm gf} =|g \rangle \langle f| + |f \rangle \langle g|$
 are Pauli matrices in the $\{\ket{\mathrm{g}}, \ket{\mathrm{f}}\}$ basis.
 and $\Omega_{\rm 2ph} = \Omega_{\rm ge} \Omega_{\rm ef}/2 \Delta$.

Experimentally, we first proceed with a modulation depth $ D=-2\pi\cross12.5$ MHz and $T = 400~\mathrm{ns}$, yielding a LZSM speed $v/2\pi  = 62.5~\mathrm{MHz}/\mu\mathrm{s}$.
A drive with larger $|D|$ would come closer to the ge- and ef- transition frequencies and therefore would couple stronger to $\ket{\rm e}$. On the other hand if $|D|$ is too small the adiabaticity is violated and we recover Rabi oscillations. Moreover, as in \cite{kuzmanovic2023high}, the robustness to frequency offsets grows with the absolute value of the modulation depth.  Therefore, as a compromise between robustness and avoiding $\ket{\rm e}$ we choose to work with $D=-2\pi\cross 12.5\mathrm{MHz}$.

\begin{figure}[h!]
	\centering
	\includegraphics[scale=0.5]{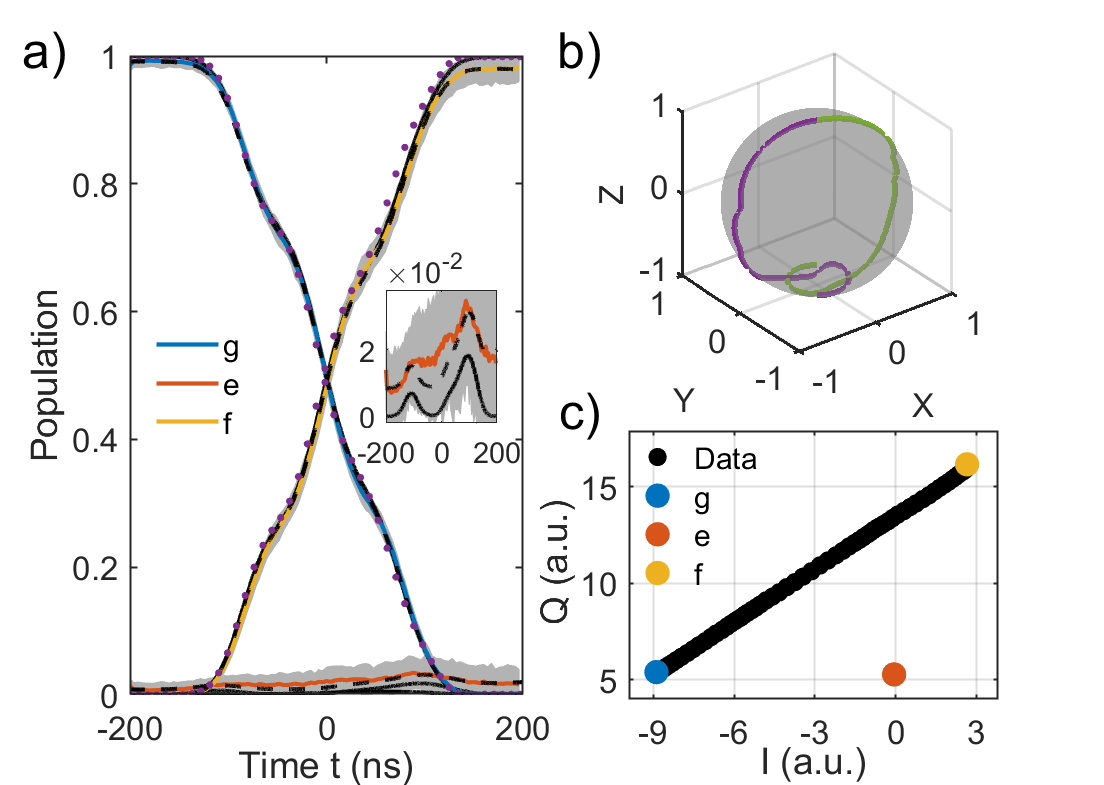}		
	\caption{(a) Population of all three states as a function of time with modulation depth $ D=-2\pi\cross12.5$ MHz. Colored solid lines are the experimental data, the grey sections are the $3\sigma$ confidence intervals, and the black solid lines are the theory based on the Hamiltonian given by Eq. \eqref{rotHamiltonian}. The black dashed lines are simulations that include finite temperature and relaxation. The purple dotted lines are the theory based on the effective Hamiltonian in Eq. \eqref{eigenstHam2}. The inset shows the population in $\ket{\rm e}$ on a smaller population-scale. (b) Majorana star representation of the results in experiment.
	(c) Experimental I- and Q-values of the readout signal: as a function of time (black), as well as the reference  ground state (blue), first excited state (red), and the second excited state (orange) responses.} 
	\label{resn5}
\end{figure}

Fig. \ref{resn5} shows the trajectory of the system in terms of the populations of the first three levels for an amplitude $\Omega/2\pi = 55.6$ MHz and frequency offset $\delta/2\pi = 0$ MHz. In fact, the same parameters were used in Fig. \ref{ieigp5} where the first instantaneous eigenstate of the Hamiltonian is followed (approximately) adiabatically from $\ket{\rm g}$ to $\ket{\rm f}$, leading to only a minimal excitation of the  $\ket{e}$ state. Experimentally we obtain $p_{\rm f} \approx 98\%$ and $p_{\rm e} < 3.48 \%$, while theoretically without relaxation and dephasing we expect a $p_{\rm f}$ to be $>99.9 \%$ and $p_{\rm e} < 1.87 \%$. Fig. \ref{resn5} also shows the evolution of the system in the Majorana constellation representation, where $|\psi \rangle = \alpha |0\rangle + \beta |1\rangle + \gamma |2\rangle$ is shown as two stars with spherical coordinates $(\theta_{1}, \varphi_{1})$ and $(\theta_{2}, \varphi_{2})$, where $\xi_{1} = \tan (\theta_{1}/2)e^{i \varphi_{1}}$ and $\xi_{2} = \tan (\theta_{2}/2)e^{i \varphi_{2}}$ are the roots of the Majorana polynomial $\mathcal{P}_{|\psi \rangle} = (\alpha/\sqrt{2}) \xi^2 - \beta \xi + \gamma /\sqrt{2} = (\alpha/\sqrt{2})( \xi - \xi_{1})(\xi - \xi_{2})$ \cite{Dogra2020}. 

\begin{figure}[h!]
	\centering
	\includegraphics[scale=0.5]{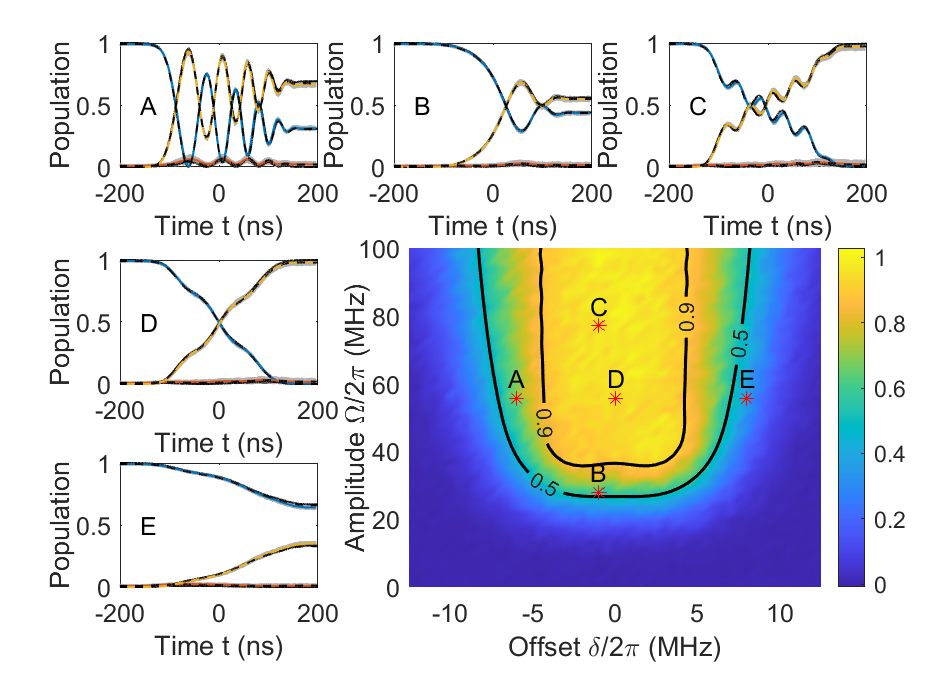}
	\caption{Population during the drive with modulation depth $-2\pi\cross12.5$ MHz. The colored lines are the experimental data,  gray regions are the three-times standard deviation, and the black solid lines are the theory based on the Hamiltonian in Eq. \eqref{rotHamiltonian}. The black dashed lines are simulations with relaxation and finite temperature. The bottom right panel shows the final $\ket{f}$ state population as a function of amplitude and detuning,  with each of the amplitude-detuning pairs from panels A-E marked. The contours show the theoretical value based on the effective Hamiltonian in Eq. \eqref{eigenstHam2}.} 
	\label{resDiffPoints}
\end{figure}

To further test the method in time domain, we measured the trajectory with randomly selected amplitude $\Omega$ and frequency offset $\delta$, seen in Fig. \ref{resDiffPoints}, while keeping the same modulation depth. 

The experimental data in figures \ref{resn5} and \ref{resDiffPoints} are supported by numerical models without any additional fitting parameters. We assume a simple Lindblad model that includes the effects of a finite temperature (73 mK effective qubit temperature, measured independently) and decay rates of the corresponding transitions given by the Fermi Rule $\Gamma_{\rm hf}/3 =  \Gamma_{\rm fe}/2 = \Gamma_{\rm eg} = 33$ kHz.

Lastly, Fig. \ref{resDiffPoints} shows the population in $\ket{f}$ as a function of the drive amplitude $\Omega$ and the frequency offset $\delta$ for a drive with the modulation depth $D=-2\pi\cross12.5\mathrm{MHz}$, demonstrating both amplitude and frequency robustness: according to the theory for $\Omega/2\pi \in [34, 62]$ MHz $\cap$ $\delta/2\pi \in [-1.0, 1.0]$ MHz we get $P_{\rm f} \geq 99 \%$. Experimentally we observe the same trend, and the theoretical $p_f=0.9$ and $p_f=0.5$ contour lines show good agreement with the experimental data.

The robustness in frequency is due to the fact that a mild offset in frequency only shifts the time at which the LZSM tunneling occurs, see Fig.1. Similarly, the robustness in amplitude is due to the fact that, once the LZSM probability is below a threshold, the effect of small fluctuations in amplitude is exponentially suppressed. In  Fig. \ref{resDiffPoints} this is observed by the formation of a plateau for the transfer probabilities.

\begin{figure}[h!]
	\centering
	\includegraphics[scale=0.52]{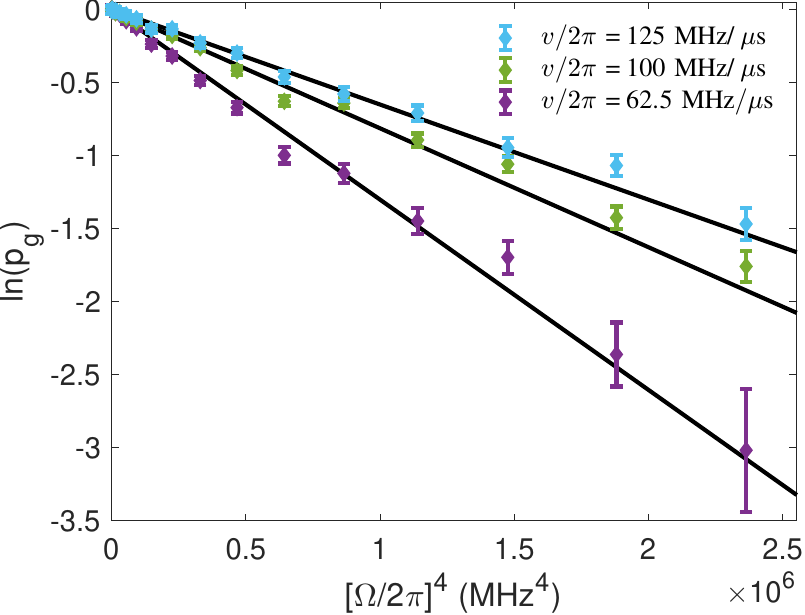}
	\caption{Natural logarithm of the population in $\ket{g}$ as a function of amplitude raised to the fourth power for a drive with modulation depths $-2\pi\cross12.5$ MHz,  $-2\pi\cross20.0$ MHz, and $-2\pi\cross25.0$ MHz. The colored diamonds are the data in the experiment and the black lines are the LZSM theory based on the effective Hamiltonian in Eq. \eqref{eigenstHam2}.}
	\label{LZ-model-fig2}
\end{figure}

Finally we study this population transfer from the point of view of the LZSM process: from the Hamiltonian given by Eq. (\ref{eigenstHam2}) we can identify the LZSM velocity and the zero-detuining gap.
Compared to the single-photon LZSM process here we get a doubling of the LZSM velocity due to the two-photon process.
In the end we obtain
\begin{equation}
	P_{\rm LZSM}^{\rm (2ph)} = \exp\left[-\pi \frac{\Omega_{\rm 2ph}^2}{4|v|}\right],
	\label{LZSM2}
\end{equation}
where $\Omega_{\rm 2ph} = \Omega^2/2\Delta$.
Even though this formula is derived for a fixed energy spectral gap, it is still approximately valid for our case since LZSM transitions happen in the cross-over region, where the amplitude reaches the maximum
value. 	Eq. (\ref{LZSM2}) also explains the amplitude robustness observed in Fig. (\ref{resDiffPoints}): indeed, small fluctuations in $\Omega$ will produce an exponentially-supressed variations of the LZSM probability.

In consequence, the probability to end up the $\ket{f}$ state is 

\begin{equation}
 	p_{\rm f} =  1- p_{\rm g} = 1-P^{\rm (2ph)}_{\rm LZSM}  
 	\label{LZSMpf}
 \end{equation}

In Fig. 4 we present the experimental data for the populations  $p_{\rm g}$ on a logarithmic scale as a function of $\Omega_{\rm 2ph}^{2} \propto \Omega^4$, in order to put clearly in evidence the linearity implied by Eq. (\ref{LZSMpf}).
The predictions of Eq. (\ref{LZSMpf}) are plotted by solid lines.

The agreement between the theory and experiments is a clear signature of the two-photon LZSM process, demonstrating the doubling on the LZSM velocity. 

In conclusion, we have observed a virtual-state LZSM process, where the transition from the ground state to the second excited state is realized by a two-photon phase-modulated pulse. We observe population transfer of 98\%, in full agreement with numerical simulations that include relaxation, and very close to the values predicted by the dynamics based only on the Hamiltonian. The population of the first excited state reaches only a few percent during this process. Moreover, we demonstrate agreement with the dynamics predicted by an effective Hamiltonian that acts on the subspace of the ground state and the second excited state, which has the form of the standard LZSM time-dependent Hamiltonian but with twice the value of single-photon LZSM speed. Besides the fundamental importance of this effect, seen as a transfer technique our approach has the advantage of robustness in frequency and amplitude: offsets in frequency only shift the time at which transitions occur, while the effect of amplitude fluctuations is exponentially suppressed.

\acknowledgments	

We are grateful to Aidar Sultanov for help with the experiments. This project has received funding from the European Union project OpenSuperQ+.
This work was performed under the Academy of Finland Centre of Excellence program (project 352925). We acknowledge the use of experimental facilities of the Low Temperature Laboratory and Micronova of the OtaNano national research infrastructure.

\bibliography{2phLZreferences}

\end{document}